\documentstyle[sprocl,psfig]{article}

\newcommand \beq{\begin{eqnarray}}
\newcommand \eeq{\end{eqnarray}}
\newcommand \be{\begin{eqnarray}}
\newcommand \ee{\end{eqnarray}}

\newcommand{\set}[2]{\newcommand{#1}{#2}}
\set{\pa}{\partial \over \partial\, }
\set{\leftvector}{\stackrel{\leftarrow}{\partial }}
\set{\rightvector}{\stackrel{\rightarrow}{\partial }}
\set{\gtrless}{\stackrel{>}{<}}

\begin{document}

\title{Nonlocal kinetic theory}

\author{Klaus Morawetz$^{1}$, V\'aclav \v Spi\v cka$^2$, Pavel Lipavsk\'y$^2$}

\address{$^1$ LPC-ISMRA, Bld Marechal Juin, 14050 Caen
 and  GANIL,
Bld Becquerel,\\ 14076 Caen Cedex 5, France\\
$^2$Institute of Physics, Academy of Sciences, Cukrovarnick\'a 10,
16200 Praha 6, Czech Republic}


\maketitle
\abstracts{
The short time behavior of a disturbed system is influenced by off-shell motion and best characterized by the reduced density matrix possessing high energetic tails. We
present analytically the formation of correlations due to collisions in an
interacting Fermionic system with and without initial correlation. 
After this short time regime the time evolution is controlled by small gradients. This leads to a nonlocal Boltzmann equation for the quasiparticle distribution and a functional relating the latter one to the reduced density matrix. The nonlocalities are presented as time and space shifts arising from gradient expansion and are leading to virial corrections in the thermodynamical limit.
}

\section{Short time regime}

The generalization of Boltzmann equation towards dense interacting systems is
a still demanding and unsolved task. A huge variety of different attempts can be found in literature to
incorporate modifications which lead to virial corrections in the equation of
state, see citations in \cite{CC90,HCB64,KL82,B69,BKKS96,SLM96}.

These kinetic equations describe different relaxation stages.
During the very fast first stage, correlations imposed by the initial
preparation of the system
are decaying \cite{BKSBKK96}. During this stage of
relaxation the quasiparticle picture is established \cite{LKKW91,MSL97a}.
After this very fast process the second state develops
during which the one-particle distribution relaxes towards the equilibrium
value with a relaxation time. During this
relaxation state the virial corrections are established and can
be consistently described by a nonlocal Boltzmann kinetic
equation \cite{SLM96}.
We will present results for both stages here.

The formation of correlations is connected with an increase of the kinetic energy or equivalently the build
up of correlation energy. This is due to rearrangement processes which let decay higher order correlation
functions until only the one - particle distribution function relaxes. Because the correlation energy is a
two - particle observable we expect that the relaxation of higher order correlations can be observed best
within this quantity. Since the total energy of the system is conserved
the kinetic energy increases on cost of the correlation energy $E_{\rm corr}(t)$ and will be calculated from kinetic equations.

The kinetic equation for the reduced density
distribution or Wigner function in Born approximation  
including memory effects but no damping
is called Levinson equation and reads in spatial
homogeneous media
\cite{MSL97a,L65,JW84}
\begin{eqnarray}\label{kinetic}
&&   \frac{\partial}{\partial t} \rho(t,k)=
2 \sum\limits_bs_a s_b
     \int \frac{dp dq}{(2 \pi)^6}
     V_{ab}(\mid q \mid )^2
\int_{t_0}^{t }d\tau 
\nonumber\\&&
\times{\rm cos}
\left [({k^2\over 2 m_a}+{p^2\over 2 m_b}-{(k-q)^2\over 2
m_a}-{(p+q)^2\over 2
m_b})(t-\tau)
\right ]
\left \{ \rho( \tau,k-q ) \rho( \tau,p+q )
\right .\nonumber\\
&& \times \left . 
(1-{\rho}( \tau,k )
-{\rho}( \tau,p ))
 - \rho( \tau,k ) \rho( \tau,p ) (1-{\rho}(
\tau,p+q )-
{\rho}( \tau,k-q )) \right \}
\end{eqnarray}
with the spin (isospin,...) degeneracy $s$.
The Wigner distribution function is normalized to the density as $s \int {d p\over (2 \pi)^3} f(p)=n$. 
The memory effect is condensed in the explicit retardation of the distribution function. This would lead to
gradient
contributions to the kinetic equation which can be shown to be responsible for the formation of high energetic
tails in the distribution function \cite{SL95}. This will be established on the second stage of relaxation and will lead to virial corrections in chapter \ref{2s}.

The second effect is contained in the energy broadening or off-shell behavior in (\ref{kinetic}). This is
exclusively related to the spectral properties of the two-particle propagator and therefore determined by
the relaxation of two-particle correlation. On this time scale the memory in the distribution functions can be neglected but
we will keep the spectral relaxation in the off-shell $\cos$-function of (\ref{kinetic}). The resulting expression for (\ref{kinetic}) describes then how two particles correlate their motion to avoid the
strong interaction regions.
This
very fast formation of the off-shell contribution to Wigner's
distribution has been found in numerical treatments of Green's
functions \cite{D841,K95}.

Starting with a sudden switching approximation we consider Coulomb
interaction and during the first transient time period the screening is
formed. This can be described by the non-Markovian Lenard - Balescu
equation \cite{Moa93} instead of the static screened equation (\ref{kinetic})
leading to the dynamical expression of the correlation energy.
To demonstrate its
results and limitations, we use Maxwellian
initial distributions with temperature $T$ neglecting degeneracy.

From (\ref{kinetic}) we find with ${\partial \over \partial t} E_{\rm corr}=-
\sum_a\int{dk\over(2\pi)^3}{k^2\over 2m_a}{\partial \over \partial t} \rho_a$
for the one-component plasma
\begin{eqnarray}
{\partial \over \partial t} {E_{\rm corr}^{\rm static}(t) \over n}&=& -{e^2 \kappa T\over 2 \hbar}{\rm Im}
\left [(1+2 z^2 ) {\rm e}^{z^2} (1- {\rm erf} (z)) -{2 z \over \sqrt{\pi}} \right ] \nonumber\\
{\partial \over \partial t} {E_{\rm corr}^{\rm dynam}(t) \over n}&=&
-{e^2 \kappa T \over  \hbar}{\rm Im}
\left [{\rm e}^{z_1^2} (1- {\rm erf} (z_1)) \right ]
\label{v1}
\end{eqnarray}
where we used $z =\omega_p \sqrt{t^2 - i t {\hbar \over T}}$ and $z_1 =\omega_p \sqrt{2 t^2 - i t {\hbar \over T}}$.
This is the analytical quantum result of the formation of correlation for statically as well as
dynamically screened potentials. 
The long time limit of (\ref{v1}) leads to the quantum Montroll correlation energy 
\cite{kker86} for the dynamical case.

\begin{figure}
\parbox[]{12.4cm}{
\parbox[]{6.1cm}{
\psfig{file=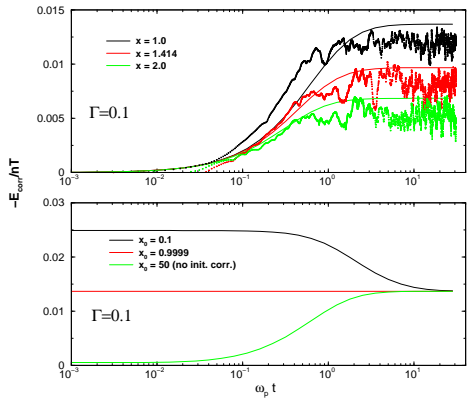,width=6.1cm,angle=-90}}
\hspace{0.2cm}
\parbox[]{5cm}{
\caption{The formation of correlation energy
$-{E}_{\rm corr}\equiv{E}_{\rm total}-{E}_{\rm init}-
{E}_{\rm coll}={E}_{\rm kin}$
in a plasma with Debye interaction $V_i$. The upper panel compares the
analytical results
(\protect\ref{state}) with MD simulations from
\protect\cite{GZ99} for
three different ratios of $\kappa_D$ to the inverse Debye length
$x=\kappa_D/\kappa$.
In the lower panel we compare theoretical predictions for the
inclusion of Debye initial correlations characterized by
$x_0=\kappa_0/\kappa$ where
$x=\kappa_D/\kappa=1$.\label{zwick3}}}
}

\end{figure}

The situation of sudden switching of interaction may be considered as an artificial one. In the simulation experiment we have initial correlations which are due to the set up within quasiperiodic boundary condition and Ewald summations.  Therefore we 
consider now initial correlations which lead  to an additional
collision term \cite{SKB99,KBKS97,D84}. This term can be written into the form of (\ref{kinetic}) with
the only difference that the occuring Wigner functions are not time
dependent but the initial ones and the interaction $V_0$ represents initial correlations\cite{MBMRK99}.
The additional collision term, ${\cal I}_0$, cancels
therefore exactly
the Levinson collision term (\ref{kinetic}) in the case that we have initially
the same interaction
as during the dynamical evolution ($V_0=V$)
and if the system starts from the
equilibrium $\rho(t)\equiv \rho_0$. We choose a model interaction of Debye potential
$V_i(q)=4 \pi e^2 / [q^2+\kappa_i^2]$
with fixed
parameter $\kappa_i=\kappa_D$
and for the initial correlations $\kappa_i=\kappa_0$ to
obtain the kinetic energy from the kinetic equation
\cite{MSL97a}
\be
E_{\rm kin}(t)=E_{\rm total}-E_{\rm init}(t)-E_{\rm coll}(t).
\label{et}
\ee
For the classical limit the time dependent collisional energy reads
\be
&&{E_{\rm coll}(t)\over nT}
=-{\sqrt{3} \Gamma^{3/2}\over 4 x}
\partial_y (y {\cal F}(y))_{y=x \tau},
\label{state}
\ee
where ${\cal F}(y)=1-{\rm e}^{y^2}{\rm erfc}(y),$
$\tau= t \omega_p/\sqrt{2}$, $x=\kappa_D/\kappa$ and
$\kappa^2=4\pi e^2 n/T=\omega_p^2 T/m$.
The plasma parameter is given as usually by
$\Gamma={e^2\over a_eT}$, where
$a_e=({3\over 4\pi n})^{1/3}$.
In Fig. \ref{zwick3}, upper panel, we compare the analytical results of
(\ref{state})
with MD simulations \cite{GZ99} using the Debye
potential $V_i$ as bare interaction.
The evolution of kinetic energy is
shown for three different ratios $x$. The agreement between theory and
simulations is quite satisfactory, in particular, the short time
behavior
for $x = 2$. The stronger initial increase of kinetic energy
observed in the simulations at $x=1$ may be due to the
finite size of the
simulation box which could more and more influence the results for
increasing range of the interaction. 
Now we include the initial correlations which leads to
\be
{E_{\rm init}(t)\over nT}
&=&-{\sqrt{3} \Gamma^{3/2}\over {2 (x_0^2-x^2)}}
\left[ x {\cal F}(x\tau) - x_0 {\cal F}(x_0 \tau)\right],
\label{dyne0}
\ee
where $x_0=\kappa_0/\kappa$ characterizing the strength of the initial
correlations with the Debye potential $V_0$ which
contains $\kappa_0$ instead of $\kappa_D$. Besides the
kinetic energy (\ref{dyne0}) from initial correlations, the total energy
$E_{\rm total}$ (\ref{et}) now includes the correlation energy at the beginning which is the total energy content of the system and can be calculated
from the long time limit of (\ref{state}) leading to
\be
{E_{\rm total}\over nT}={\sqrt{3} \Gamma^{3/2}\over 2 (x+x_0)}.
\ee
The result (\ref{et}) is seen in Fig. \ref{zwick3}, lower panel. We
observe that if the initial correlation is characterized by a potential
range
larger than the Debye screening length, $x_0 < 1$, the initial state is
over--correlated, and the correlation energy starts at a higher absolute
value than without initial correlations relaxing towards the correct
equilibrium value.
If, instead, $x_0 = 1$ no change of correlation energy is observed, as expected.
Similar trends have been observed in numerical solutions
\cite{SKB99}.

The characteristic time of formation of correlations at high temperature
limit is given by the inverse plasma frequency
$\tau_c\approx{1\over\omega_p}={\sqrt{2}\over v_{\rm th}\kappa}$.
This confirms the
numerical finding of \cite{BKSBKK96} that the correlation or memory
time is proportional to the range of interaction.
In the low temperature region, i.e., in a highly degenerated system
$\mu\gg T$, one finds \cite{MSL97a} that the build up time
is the inverse Fermi energy, $\hbar/\epsilon_f$ in agreement with the quasiparticle formation
time known as Landau's criterion.
Indeed, the
quasiparticle formation and the build up of correlations are two
alternative views of the same phenomenon.

\section{Virial regime}\label{2s}

Now we proceed and investigate how the relaxation into equilibrium is performed by the system. We will convince ourselves firstly that the direct use of reduced density matrix as done so far for short time regimes is not possible.

On larger time scales we expect smoothed time-space gradients and 
expand the Levinson equation
 for the reduced density matrix up to first order memory 
to obtained from (\ref{kinetic})
\beq
{\partial
    \over \partial t}\rho={\cal I}_{\rm Boltz}+{\partial
    \over \partial t}{\cal I}_{\rm corr}
\label{redc}
    \eeq
where besides the usual Boltzmann term, ${\cal I}_{\rm Boltz}$, there appears an
additional off-shell term, ${\cal I}_{\rm corr}$, which describes the correlation in the
reduced density matrix. For the density balance we see from
(\ref{redc}) that the correlation part from the right side $\int dp {\cal I}_{\rm
corr}=n_c$ which is known as the Beth-Uhlenbeck equation of state 
\cite{BU37,SZ79,SR87}
cancels exactly with the left hand side of (\ref{redc}) according to the definition $\int dp \rho =n_f+n_c$. So we 
have to face the exact cancelation of $n_c$ in the balance
equation and the incorrect conservation of $n_f$. We will show that the virial corrections and the correlated density will appear instead from {\it internal} gradients.
That the correlated part ${\cal I}_{\rm
corr}$ does not provide the virial corrections but a relation between reduced density matrix and quasiparticle distribution has not been noted in the literature.

The second puzzle is a more obvious hint that the Levinson
equation fails to describe the long time evolution of a system. We can iterate the time derivative inside the
collision integral on the right hand side of (\ref{redc}) to create an infinite series. Neglecting
backscattering terms we estimate
\begin{eqnarray}
\frac{\partial}{\partial
t}f \ge {\zeta [f]} \, I_{Boltz} [f]
\end{eqnarray}
 with
${\zeta}= \sum\limits_{n=0}^{\infty} {\eta}^{2^n-1}$
and $\eta$ is similar to the backflow (second order response) \cite{SL95}
and is a positive number but not every time smaller than
one. Therefore $\zeta$ appears to be divergent at certain
momentum situations. Therefore we conclude that the gradient
expanded Levinson equation leads not to a stationary solution in any case. 
This has also been observed
in numerical solutions of these non-Markovian kinetic equations, see
figure 4 of 
\cite{KBKS97}.
 We will show that a proper
balance between off-shell terms
in the reduced density and the kinetic equation for the pole part
is necessary to ensure stationary solutions.

The approach presented now is based on the real-time Greenfunction
technique. We consider the two independent correlation functions for
Fermionic creation operators
$G^> (1,2) =   <a(1) a^+(2)>$ and
$G^< (1,2) =  <a^+(2) a(1)>$,
where cumulative variables means time, space, spin, etc $1=t,x,s...$.
The time diagonal part of $G^<$ yields the reduced density matrix
according to $  \rho(x_1,x_2,t)=G^<(1,2)_{t_1=t_2}$.
For the latter one the time diagonal Kadanoff and Baym equation of motion
 reads \cite{KB62}

\begin{eqnarray}
&&  i [G_0^{-1},\rho](x_1,x_2,t)=
 \int\limits_{t_0}^{t} d t'\int dx'
\left [
  G^<(x_1,x',t,t') \Sigma^>(x',x_2,t',t) \right .\nonumber\\
 &&\left . +
\Sigma^>(x_1,x',t,t') G^<(x',x_2,t',t)
-G^>(x_1,x',t,t')\Sigma^<(x',x_2,t',t)
\right .
\nonumber\\&&
\left . -
\Sigma^<(x_1,x',t,t') G^>(x',x_2,t',t)
 \right ]
\label{levin}
\end{eqnarray}
with the Hartree- Fock drift term $G_0^{-1}$.
The right hand side contains a non-Markovian collision integral and can be considered as a precursor of Levinson equation (\ref{kinetic}).
Using gradient approximation we obtain a kinetic equation of the structure (\ref{redc}) and see that the time derivative of the Wigner function on the
left side of (\ref{levin}) combines with the time derivative of the
off-shell part of right hand side into a time derivative of a quasiparticle
distribution $f$
\beq
\rho\approx z f-\int {d \omega \over 2 \pi} {{\cal P}'\over \omega -\epsilon}\sigma^<(\omega)
\label{an2}
\eeq
with ${\cal P}'$ the derivative of the principal value.
The wave function renormalization has been abbreviated as
$z=(1-\partial_{\omega}\sigma)^{-1}\approx 1+\partial_{\omega}\sigma$ which can be confirmed 
using the spectral decomposition
$\sigma(\omega)=\int {d \omega'
\over 2 \pi} {P \over \omega-\omega'} \gamma (\omega')$
with the imaginary part $\gamma=\sigma^>+\sigma^<$.
We have fulfilled the task and give
with (\ref{an2}) a connection between reduced density matrix $\rho$
and the quasiparticle distribution $f$. We extrapolate the ansatz also for the correlation functions 
\beq\label{ansatz2}
g^{\gtrless}(\omega)&=&2 \pi z \delta(\omega-\epsilon) \left (\matrix{1-f\cr f}\right )-{P'\over \omega-\epsilon}\sigma^{\gtrless}(\omega).
\eeq
The spectral identity $a=g^>+g^<$ proofs that this ansatz is consistent with the extended
quasiparticle picture which is obtained 
for small imaginary parts of the self-energy $\gamma$ \cite{SL95,kker86,SZ79,SR87,C66a,BD68,KKL84,KM93}
\begin{eqnarray}\label{quasi}
a(p\omega rt)&=&{2\pi\delta(\omega - \epsilon(prt)) \over
1+\frac{\partial \sigma(p,\omega,r,t)}{\partial \omega}|_{\omega =\epsilon}}
- \gamma(p,\omega,r,t)\frac{\partial}{\partial \omega} \frac{{\cal P}}{
\omega-\epsilon(prt)}.
\end{eqnarray}
The quasiparticle energies $\epsilon(prt)$ are a solution of the
dispersion relation
$\omega-\frac{p^2}{2m}-\sigma(p\omega rt)=0$.
It is noteworthy to remark that (\ref{quasi}) fulfills the
spectral sum rule \cite{KKL84}
and the energy weighted sum rule \cite{SLM96}.
The limit of small scattering rates has been first introduced by Craig
\cite{C66a}. An inverse functional $f[\rho]$ has been constructed in \cite{BD68}. For equilibrium nonideal plasmas this approximation has been employed by \cite{SZ79,KKL84} and under the name of the {\it generalized
Beth-Uhlenbeck approach} has been used by \cite{SR87} in
nuclear matter for studies of the correlated density. The authors in \cite{KM93}
have used this approximation with the name
{\it extended quasiparticle approximation} for the study of the mean removal energy and
high-momenta tails of Wigner's distribution. The non-equilibrium form has been
derived finally as {\it modified Kadanoff and Baym ansatz} \cite{SL95}.
Using this ansatz (\ref{ansatz2}) from the Kadanoff-Baym equations (\ref{levin})
the known 
Landau-Silin equation for the quasiparticle distribution $f$ follows
  \begin{equation}
{\partial \over \partial t} f + \nabla_k \epsilon \nabla_r f - \nabla_r
  \epsilon \nabla_k f =z \left ((1-f)\sigma^<-f\sigma^> \right ).
  \label{Silin}
\end{equation}
We repeat that the Landau-Silin equation (\ref{Silin}) is coupled with a
functional that specifies a relation between the quasiparticle
distribution $f$ and Wigner's function $\rho$ via (\ref{an2}).
Using different approximations for the self energy we obtain all known kinds
of kinetic equations with the generalization that the internal
gradients of collision integrals will yield the nonlocal or virial corrections.

The quality of the extended quasiparticle approximation (\ref{quasi}) can
be seen in the figure of the contribution by P. Lipavsky.  One can 
see that the off-shell contribution given by the difference between the
Wigner and the Fermi-Dirac distributions is not small, in particular at 
the high momenta region where the power-law off-shell tails always 
dominate over the exponentially falling quasiparticle distribution. 
Formula (\ref{quasi}) provides inevitable and sufficiently precise off-shell 
corrections.

In order to show the nonlocal character of the kinetic equation (\ref{Silin}) explicitly, we choose as an approximation for the self energy the ladder or ${\cal T}$-matrix
approximation appropriate for dense interacting systems of short range potentials
\begin{eqnarray}
\Sigma^<(1,2)
&=&
{\cal T}^R(1,\bar 3;\bar 5,\bar 6)
{\cal T}^A(\bar 7,\bar 8;2,\bar 4)
G^>(\bar 4,\bar 3)
G^<(\bar 5,\bar 7)
G^<(\bar 6,\bar 8).
\label{tr3}
\end{eqnarray}
For non-degenerate systems, the gradient expansion has been carried through by
B{\"a}rwinkel \cite{B69}. One can see in B{\"a}rwinkel's papers, that the
scattering integral is troubled by a large set of gradient corrections.
This formal complexity seems to be the main reason why most authors
either neglect gradient corrections at all \cite{D84,bm90} or provide
them buried in multi-dimensional integrals \cite{MR95a,L90}.
For a degenerate system, the set of gradient
corrections to the scattering integral is even larger than for rare
gases studied by B{\"a}rwinkel, see \cite{SLM96}.

The quasiclassical limit with all linear gradients kept
is a tedious but straightforward algebraic exercise.  It
results in one nongradient and nineteen gradient terms that are
analogous to those found within the chemical physics \cite{NTL91,H90}.
All these terms can be recollected into a nonlocal and noninstantaneous
scattering integral that has an intuitively appealing structure of a
nonlocal Boltzmann equation \cite{SLM96} 
\begin{eqnarray}
\!\!\!&&{\partial f_1\over\partial t}+{\partial\varepsilon_1\over\partial k}
{\partial f_1\over\partial r}-{\partial\varepsilon_1\over\partial r}
{\partial f_1\over\partial k}
=\sum_b\int{dpdq\over(2\pi)^5}\delta\left(\varepsilon_1+
\varepsilon_2-\varepsilon_3-\varepsilon_4+2\Delta_E\right)
z_1z_2z_3z_4
\nonumber\\
\!\!\!&&\times 
\left|{\cal T}_{ab}\!\left(\varepsilon_1\!+\!\varepsilon_2\!-\!
\Delta_E,k\!-\!{\Delta_K\over 2},p\!-\!{\Delta_K\over 2},
q,r\!-\!\Delta_r,t\!-\!{\Delta_t\over 2}\!\right)\right|^2
\Biggl(\!1\!-\!{1\over 2}{\partial\Delta_2\over\partial r}\!-\!
{\partial\bar\varepsilon_2\over\partial r}
{\partial\Delta_2\over\partial\omega}\!\Biggr)
\nonumber\\
\!\!\!&&\times
\Bigl[f_3f_4\bigl(1-f_1\bigr)\bigl(1-f_2\bigr)-
\bigl(1-f_3\bigr)\bigl(1-f_4\bigr)f_1f_2\Bigr],
\label{9}
\end{eqnarray}
with Enskog-type shifts of arguments \cite{SLM96}:
$f_1\equiv f_a(k,r,t)$, $f_2\equiv f_b(p,r\!-\!\Delta_2,t)$,
$f_3\equiv f_a(k\!-\!q\!-\!\Delta_K,r\!-\!\Delta_3,t\!-\!\Delta_t)$, and
$f_4\equiv f_b(p\!+\!q\!-\!\Delta_K,r\!-\!\Delta_4,t\!-\!\Delta_t)$.
In agreement with \cite{NTL91,H90}, all gradient corrections result
proportional to derivatives of the scattering phase shift
\mbox{$\phi={\rm Im\ ln}{\cal T}^R_{ab}(\Omega,k,p,q,t,r)$},
\begin{equation}
\begin{array}{lclrclrcl}
\Delta_2&=&
{\displaystyle\left({\partial\phi\over\partial p}-
{\partial\phi\over\partial q}-{\partial\phi\over\partial k}
\right)_{\varepsilon_3+\varepsilon_4}}&\ \
\Delta_3&=&
{\displaystyle\left.-{\partial\phi\over\partial k}
\right|_{\varepsilon_3+\varepsilon_4}}&\ \
\Delta_4&=&
{\displaystyle-\left({\partial\phi\over\partial k}+
{\partial\phi\over\partial q}\right)_{\varepsilon_3+\varepsilon_4}}
\\ &&&&&&&&\\
\Delta_t&=&
{\displaystyle \left.{\partial\phi\over\partial\Omega}
\right|_{\varepsilon_3+\varepsilon_4}}&\ \
\Delta_E&=&
{\displaystyle \left.-{1\over 2}{\partial\phi\over\partial t}
\right|_{\varepsilon_3+\varepsilon_4}}&\ \
\Delta_K&=&
{\displaystyle \left.{1\over 2}{\partial\phi\over\partial r}
\right|_{\varepsilon_3+\varepsilon_4}},
\end{array}
\label{8}
\end{equation}
and $\Delta_r={1\over 4}(\Delta_2+\Delta_3+\Delta_4)$. After derivatives, 
$\Delta$'s are evaluated at the energy shell $\Omega\to\varepsilon_3+
\varepsilon_4$.

The $\Delta$'s in arguments of distribution
functions in (\ref{9}) remind non-instant and non-local
corrections in the scattering-in integral for classical particles.
The displacements of the asymptotic states are given
by $\Delta_{2,3,4}$. The time delay enters in an equal way the
asymptotic states 3 and 4. The momentum gain $\Delta_K$ also appears
only in states 3 and 4. Finally, there is the energy gain which is discussed in \cite{LSM99}. 
We remind that the non-localities should be viewed as
a compact form of gradient corrections.

Despite its complicated form it is possible to solve this kinetic equation with standard Boltzmann numerical codes and to implement the shifts \cite{MLSCN98}. Therefore we have calculated the shifts for different realistic nuclear potentials \cite{MLSK98}. The numerical solution of the nonlocal kinetic equation has shown an observable effect in the dynamical particle spectra of around $10\%$. The high energetic tails of the spectrum is enhanced due to more violent collisions in the early phase of nuclear collision. Therefore the nonlocal corrections lead to an enhanced production of preequilibrium high energetic particles.

\section{Thermodynamic properties}

The meaning of nonlocal shifts can be best seen on 
thermodynamic observables like density $n_a$ of 
particles $a$, density of energy $\cal E$, and the stress tensor 
${\cal J}_{ij}$ which conserve within the nonlocal and noninstantaneous
kinetic equation (\ref{9}). Integrating (\ref{9}) over momentum $k$ with 
factors $\varepsilon_1$, $k$ and unity one finds \cite{SLM96} that 
each observable has the standard quasiparticle part following from the 
drift
\begin{eqnarray}
{\cal E}^{\rm qp}&=&\sum_a\int{dk\over(2\pi)^3}{k^2\over 2m}f_1+
{1\over 2}\sum_{a,b}\int{dkdp\over(2\pi)^6}
T_{ab}(\varepsilon_1+\varepsilon_2,k,p,0)f_1f_2,
\nonumber\\
{\cal J}_{ij}^{\rm qp}&=&\sum_a\int{dk\over(2\pi)^3}\left(k_j
{\partial\varepsilon_1\over\partial k_i}+
\delta_{ij}\varepsilon_1\right)f_1-\delta_{ij}{\cal E}^{\rm qp},
\nonumber\\
n_a^{\rm qp}&=&\int{dk\over(2\pi)^3}f_1,
\label{10a}
\end{eqnarray}
and the $\Delta$-contribution following from the scattering integral
\begin{eqnarray}
\Delta {\cal E}&=&{1\over 2}\sum_{a,b}\int{dkdpdq\over(2\pi)^9} P
(\varepsilon_1+\varepsilon_2)\Delta_t,
\nonumber\\
\Delta {\cal J}_{ij}&=&{1\over 2}
\sum_{a,b}\int{dkdpdq\over(2\pi)^9} P
\left[(p\!+\!q)_i\Delta_{4j}+(k\!-\!q)_i\Delta_{3j}-p_i\Delta_{2j}\right],
\nonumber\\
\Delta n_a&=&\sum_b\int{dkdpdq\over(2\pi)^9} P \Delta_t,
\label{10}
\end{eqnarray}
where $P=|{\cal T}_{ab}|^22\pi\delta(\varepsilon_1\!+\!\varepsilon_2\!-
\!\varepsilon_3\!-\!\varepsilon_4)f_1f_2(1\!-\!f_3\!-\!f_4)$. 

The density of energy ${\cal E}={\cal E}^{\rm qp}+\Delta{\cal E}$
alternatively results from Kadanoff and Baym formula, ${\cal E}=
\sum_a\int{dk\over(2\pi)^3}\int{d\omega\over 2\pi}
{1\over 2}\left(\omega+{k^2\over 2m}\right)G^<_a(\omega,k,r,t)$,
with $G^<$ in the extended quasiparticle approximation (\ref{ansatz2}).
Its complicated form, however, shows that $\cal E$ cannot be easily
inferred from an eventual experimental fit of the kinetic equation as it
has been attempted in \cite{SG86}. The conservation of ${\cal E}$ 
generalizes the result of Bornath, Kremp, Kraeft and Schlanges 
\cite{BKKS96} restricted to non-degenerated systems. The particle 
density $n_a=n_a^{\rm qp}+\Delta n_a$ is also 
obtained from (\ref{an2}) via the definition, $n_a=\int{d\omega\over 2\pi}
{dk\over(2\pi)^3}G^<$. This confirms that the extended quasiparticle 
approximation is thermodynamically consistent with the nonlocal and
noninstantaneous corrections to the scattering integral.

For equilibrium distributions, formulas (\ref{10a}) and (\ref{10})
provide equations of state. Two known cases are worth to compare.
First, the particle density $n_a=n_a^{\rm qp}+\Delta n_a$ is identical
to the quantum Beth-Uhlenbeck equation of state \cite{SR87,BKKS96},
where $n_a^{\rm qp}$ is called the free density and $\Delta n_a$ the
correlated density. Second, the virial correction to the stress tensor
has a form of the collision flux contribution known in the theory of
moderately dense gases \cite{CC90,HCB64}.

\section{Summary}

While the short time behavior is described by off-shell transport
condensed in the reduced density matrix, the later stage of evolution
is described by a nonlocal Boltzmann equation. This equation is
derived in the quasiclassical limit and leads to consistent
thermodynamics in equilibrium including binary correlations. The
latter ones represent the second virial coefficients in the low
density limit. The presented kinetic equation unifies achievements of
dense gases and the kinetic theory of quasiparticle transport in
quantum systems. The equation has been shown applicable in recent
simulation codes.

\end{document}